# THE GALACTIC FOREGROUNDS ANGULAR SPECTRA

*A.V. Chepurnov*
Special Astrophysical Observatory
Nizhnij Arkhyz, Karachai-Cherkess Republic  357147 Russia
*chepurnov@gmx.de*

**Abstract**
Galactic synchrotron and free-free foregrounds angular spectra are analytically estimated with account for interstellar turbulence and radiating process physics. Unknown parameters of the spectra are obtained by fitting to observational data.

## Introduction

The problem of Galactic screening foregrounds is of an exclusive importance for CMB anisotropy experiments aiming to obtain basic cosmological parameters by an accurate measurement of Sakharov oscillations [17,18,16]. The information about Galactic continuous emission is used to determine radio frequency – angular scale region where CMB fluctuations dominate.

Present estimations of the most unclear aspect of this problem, angular spectra, are based on direct calculation from observational data and empirical two-parametrical (amplitude — angular spectrum index) spectrum model [3]. However, applying this approach we encounter the lack of observational data and their low quality (for example, see [9]), which can make direct angular spectrum estimations doubtful.

So it seems convenient to have a parametrization that accounts for a radiating process physics. In this case we can incorporate an essentially heterogeneous information, what can compensate primary data imperfections. In addition, as fitting parameters may have a physical meaning in this approach, we can find some new information about interstellar media.

Working in this direction we didn't find in existing publications a convenient analytical method allowing to bind physical parameters of interstellar media with statistical properties of observed radio images, so a considerable part of this paper is devoted to the development of such a method.

## 1. Spectral properties of the Galactic radio emission

### *1.1. Basic terms*

In this work we have to deal with different values and transforms involving stationary processes. The most convenient and in some cases the only way to obtain a desired result is to use a random process spectral representation [28, 15]:

$$s(\vec{r}) = \int e^{i\vec{k}\vec{r}}\, \Phi(d\vec{k}). \tag{1.1.1}$$

where $\Phi$ is a stochastic spectral measure of a stationary process $s$.

This integral can be interpreted as a sum of flat waves, which complex amplitudes are determined by a stochastic spectral measure of a wave vector space elementary volume, while these measure elements obey the following symbolic rule:

$$\overline{\Phi(d\vec{k})\Phi^*(d\vec{k}')} = \delta_{\vec{k},\vec{k}'}\, F^2(\vec{k})\, d\vec{k}, \tag{1.1.2}$$

where

$$\delta_{\vec{k},\vec{k}'} = \begin{cases} 0, & \vec{k} \neq \vec{k}' \\ 1, & \vec{k} = \vec{k}' \end{cases}, \tag{1.1.3}$$

and $F^2(\vec{k})$ is a power spectrum of a process $s$:
$$\overline{s(\vec{0})\,s(\vec{r})} = \int e^{i\vec{k}\vec{r}} F^2(\vec{k})\,d\vec{k}. \qquad (1.1.4)$$

It is possible to employ an alternative *symbolic* form of the spectral representation, which is completely equivalent to the previous one:
$$s(\vec{r}) = \int e^{i\vec{k}\vec{r}} F(\vec{k})\,\xi(\vec{k})\sqrt{d\vec{k}}, \qquad (1.1.5)$$
$$\overline{\xi(\vec{k})\,\xi^*(\vec{k}')} = \delta_{\vec{k},\vec{k}'}. \qquad (1.1.6)$$

We use the latter here despite of the odious $\sqrt{d\vec{k}}$, as the practice shows it to be more convenient.

We also assume the following to be correct for real random processes:
$$\xi(-\vec{k}) = \xi^*(\vec{k}). \qquad (1.1.7)$$
$$\overline{\xi^2(\vec{k})} = 0. \qquad (1.1.8)$$

## 1.2. Angular spectrum through the spatial one

We consider a model when an observer is located in the center of some sphere filled with luminous medium, and a radiation source function $s(\vec{r})$ is assumed to be a stationary random process with the known power spectrum. The sphere is cut from infinite space by some known function $w(r)$ of distance $r$ from the sphere's center, referred to as *weighting function*. (This model is adequate enough if respective variation of a distance to radiating medium edge is considerably less than unity at considered angular scales).

We are interested here in the angular spectrum of an observed picture, given by the following function of angular coordinates:
$$S(\theta,\varphi) \equiv \int_0^\infty w(r)\,s(r,\theta,\varphi)\,dr. \qquad (1.2.1)$$

The source function will be substituted below with its spectral representation:
$$s(\vec{r}) = \int e^{i\vec{k}\vec{r}} F(\vec{k})\,\xi(\vec{k})\sqrt{d\vec{k}}.$$

Let the original random process to be isotropic. If so, accounting for
$$\vec{k}\,\vec{r} = r\cdot(k_x \sin\theta\cos\varphi + k_y \sin\theta\sin\varphi + k_z \cos\theta)$$
we have the following expression for an angular autocorrelation function:
$$C(\theta) \equiv \overline{S(0,0)\cdot S^*(\theta,0)} =$$
$$= \int_0^\infty w(r)\,dr \cdot \int_0^\infty w(r')\,dr' \cdot \int F(k)\sqrt{d\vec{k}} \cdot \int F(k')\sqrt{d\vec{k}'}\cdot e^{i(-k_x'r'\sin\theta - k_z'r'\cos\theta + k_z r)} \cdot \overline{\xi(\vec{k})\,\xi^*(\vec{k}')}$$

Having evaluated the averaging we derive:
$$C(\theta) = \int_0^\infty w(r)\,dr \cdot \int_0^\infty w(r')\,dr' \cdot \int F^2(k)\,d\vec{k} \cdot e^{i(-k_x'r'\sin\theta - k_z r'\cos\theta + k_z r)}$$

After integration over wave vector directions, we have the following expression:
$$C(\theta) = 4\pi \int_0^\infty w(r)\,dr \cdot \int_0^\infty w(r')\,dr' \cdot \int_0^\infty k^2\,dk \cdot F^2(k) \cdot \frac{\sin k\sqrt{r^2 - 2rr'\cos\theta + r'^2}}{k\sqrt{r^2 - 2rr'\cos\theta + r'^2}}.$$

With the account for
$$C_l^2 \equiv 2\pi \int_0^\pi C(\theta)\,P_l(\cos\theta)\sin\theta\,d\theta. \qquad (1.2.2)$$
we finally obtain the desired angular spectrum:

$$C_l^2 = 16\pi^2 \int_0^\infty k^2 dk \cdot F^2(k) \cdot \left( \int_0^\infty j_n(kr) w(r) dr \right)^2. \tag{1.2.3}$$

We can see, that the expression of angular spectrum through the spatial one takes the form of an integral transform with kernel being dependent on the weighting function.

However, this kernel isn't convenient for direct computation because of an oscillating function (spherical Bessel's function $j_n(kr)$) in the integral. So it may be useful to find some direct analytical expression for it, with given adequate enough weighting function.

### 1.3. Kernel approximation

Let $w(\vec{r})$ to be Gaussian:

$$w(r) = e^{-\frac{r^2}{R^2}}, \tag{1.3.1}$$

where $R$ is a distance from a luminous region edge.

Then the transform may be written in the following form:

$$C_l^2 = \frac{16\pi^2}{R} \cdot \int_0^\infty u^2 du \cdot F^2(u/R) \cdot Q_l(u), \tag{1.3.2}$$

where

$$Q_l(u) \equiv \left( \int_0^\infty j_l(uv) e^{-v^2} dv \right)^2. \tag{1.3.3}$$

Using an integral representation of the hypergeometric function,

$$_1F_1(a; c; -1/z) = \frac{\Gamma(c)}{\Gamma(a)} \cdot z^a \cdot \int_0^\infty e^{-zt} \cdot t^{a-\frac{c+1}{2}} \cdot J_{c-1}(2\sqrt{t}) dt,$$

we can find

$$Q_l(u) = \left( \frac{\sqrt{\pi}}{2u} \cdot \frac{\Gamma\left(\frac{l+1}{2}\right)}{\Gamma\left(\frac{2l+3}{2}\right)} \cdot \left(\frac{u^2}{4}\right)^{\frac{l+1}{2}} \cdot _1F_1\left(\frac{l+1}{2}; \frac{2l+3}{2}; -\frac{u^2}{4}\right) \right)^2.$$

On the other hand, using an alternative integral form of $_1F_1$,

$$_1F_1(a; c; z) = \frac{\Gamma(c)}{\Gamma(a)\Gamma(c-a)} \cdot \int_0^1 e^{zt} \cdot t^{a-1} \cdot (1-t)^{c-a-1} dt,$$

we have:

$$Q_l(u) = \left( \frac{\sqrt{\pi}}{2u} \cdot \frac{\left(\frac{u^2}{4}\right)^{\frac{l+1}{2}}}{\Gamma\left(\frac{l+2}{2}\right)} \cdot \int_0^1 e^{-\frac{u^2}{4}t} \cdot t^{\frac{l-1}{2}} \cdot (1-t)^{\frac{l}{2}} dt \right)^2.$$

Let us find an approximation of this expression for large $l$ values. Let us denote

$$J \equiv \left(\frac{u^2}{4}\right)^{\frac{l+1}{2}} \cdot \int_0^1 e^{-\frac{u^2}{4}t} \cdot t^{\frac{l-1}{2}} \cdot (1-t)^{\frac{l}{2}} dt = \int_0^{u^2/4} e^{-z} \cdot z^{\frac{l-1}{2}} \cdot \left(1 - \frac{4z}{u^2}\right)^{\frac{l}{2}} dz.$$

Let $\alpha = \frac{u}{l}$ to be constant:

$$J = \int_0^{\alpha^2 l^2/4} e^{-z} \cdot z^{\frac{l-1}{2}} \cdot \left(1 - \frac{4z}{\alpha^2 l^2}\right)^{\frac{l}{2}} dz.$$

The integrand of the latter expression is positive and has a single maximum at $z_0 \approx l/2$ while $l \gg 1$, and with this condition the bulk of the area is located near the maximum. Since the maximum's form is affected only by the first two terms, the third one can be factored out of the integral with its value at $z_0$. Taking into account fast decrease of a remaining integrand, we can spread upper integration limit to infinity. So we have:

$$J \approx \left(1 - \tfrac{2}{\alpha^2 l}\right)^{\tfrac{l}{2}} \cdot \int_0^\infty e^{-z} \cdot z^{\tfrac{l-1}{2}} dz = \left(1 - \tfrac{2}{\alpha^2 l}\right)^{\tfrac{l}{2}} \cdot \Gamma\!\left(\tfrac{l+1}{2}\right) \approx \Gamma\!\left(\tfrac{l+1}{2}\right) \cdot e^{-\tfrac{1}{\alpha^2}},$$

$$Q_l(u) \approx \frac{\pi}{4u^2}\left(\frac{\Gamma\!\left(\tfrac{l+1}{2}\right)}{\Gamma\!\left(\tfrac{l+2}{2}\right)}\right)^2 \cdot e^{-2\tfrac{l^2}{u^2}}.$$

Using the gamma-function asymptotic,

$$\Gamma(z) \approx \sqrt{\tfrac{2\pi}{z}} \cdot e^{-z} \cdot z^z, \quad z \gg 1,$$

we finally obtain the following approximation of the kernel:

$$Q_l(u) \approx \frac{\pi}{2lu^2} \cdot e^{-2\tfrac{l^2}{u^2}}, \quad l \gg 1. \qquad (1.3.4)$$

This approximation works with appropriate precision for $l > 15$ uniformly over all 3-d wave vector values, so it can be used in the whole range where the original model is correct.

The following sections contain some special cases, important for practical applications.

### 1.4. Power-law 3-d spectrum

If a source function has an unlimited power-law spectrum, the respective sphere projection has also a power-law spectrum, and the spectral indexes coincide:

$$F^2(k) = \frac{F_0^2}{k^\alpha}, \qquad (1.4.1)$$

$$C_l^2 = \frac{4\pi^3 \, \Gamma\!\left(\tfrac{\alpha-1}{2}\right)}{2^{\tfrac{\alpha-1}{2}}} \cdot \frac{F_0^2 \cdot R^{\alpha-1}}{l^\alpha}. \qquad (1.4.2)$$

Let us introduce some cut-off scale ("outer scale") $L$ into the initial 3-d spectrum:

$$F^2(k) = \frac{F_0^2}{k^\alpha} \cdot e^{-\left(\tfrac{2\pi}{kL}\right)^2}. \qquad (1.4.3)$$

In this case the respective angular spectrum changes its spectral index from 1 for low $l$ values to the spectral index of the source function for large $l$'s in the vicinity of $l_0 = 2\pi R/L$:

$$C_l^2 = \frac{4\pi^3 \, \Gamma\!\left(\tfrac{\alpha-1}{2}\right)}{2^{\tfrac{\alpha-1}{2}}} \cdot \frac{F_0^2 \cdot R^{\alpha-1}}{l^\alpha \cdot \left(1 + \tfrac{1}{2}\left(\tfrac{2\pi R}{l \cdot L}\right)^2\right)^{\tfrac{\alpha-1}{2}}}. \qquad (1.4.4)$$

A remarkable property of the spectrum (1.4.4) lies in the fact, that opposite to the spectrum (1.4.2) an amplitude of a harmonic with the fixed $l$ is saturated when $R \to \infty$, which is caused by the influence of the outer scale $L$. Indeed, let us fix some angular scale $\beta$. Then at the distance more than $L/\beta$ there will be no fluctuations, exceeding angle $\beta$, and the increase of distance $R$ to the border of the radiating region over this limit will not cause the increase of the amplitude of an angular harmonic with the typical scale $\beta$.

Decrease of the spectral index when going from small angular scales (large $l$ values) to the big ones (small $l$ values) is explained by the same effect. Until the distance $L/\beta$ doesn't stop exceeding $R$, the change (growth) of the angular harmonic with the increase of scale $\beta$ occurs exclusively at the account of the increase of the amplitudes of spatial harmonics of the original random process at every distance inside the sphere of radius $R$ (according to (1.4.3)). However, beginning with the angular scale $\beta_0 = L/R$, with further increase of $\beta$ the "effective sphere" of radius $L/\beta$ (i.e. the sphere, where the amplitude of the angular harmonic is formed) shrinks, which causes slowing down of the increase of the respective angular harmonic amplitude.

The reason causing the unity spectral index at low $l$'s is directly seen from the expression (1.4.2): if with $l$ decrease (increase of $\beta$) we will on purpose decrease the distance $R$ proportionally to $l$, imitating shrinking of the "effective sphere", then the spectrum will behave as $l^{-1}$.

## 1.5. The case of squared Kolmogorov process

Let a source function to be a squared stationary random process with known properties:
$$s(\vec{r}) = (\overline{S} + \Delta S(\vec{r}))^2, \qquad (1.5.1)$$
$$\Delta S(\vec{r}) = \int e^{i\vec{k}\vec{r}} F(\vec{k}) \xi(\vec{k}) \sqrt{d\vec{k}}.$$

Respective autocorrelation function is equal to
$$C(\vec{r}) = \overline{s(\vec{r})s(\vec{0})} =$$
$$= \overline{\left(\overline{S}^2 + 2\overline{S}\cdot\int\sqrt{d\vec{k}_1}e^{i\vec{k}\vec{r}}F(\vec{k}_1)\cdot\xi(\vec{k}_1) + \int\sqrt{d\vec{k}_2}e^{i\vec{k}\vec{r}}F(\vec{k}_2)\cdot\xi(\vec{k}_2)\cdot\int\sqrt{d\vec{k}_3}e^{i\vec{k}\vec{r}}F(\vec{k}_3)\cdot\xi(\vec{k}_3)\right)}\cdot$$
$$\cdot\overline{\left(\overline{S}^2 + 2\overline{S}\cdot\int\sqrt{d\vec{k}_1'}F(\vec{k}_1')\cdot\xi(\vec{k}_1') + \int\sqrt{d\vec{k}_2'}F(\vec{k}_2')\cdot\xi(\vec{k}_2')\cdot\int\sqrt{d\vec{k}_3'}F(\vec{k}_3')\cdot\xi(\vec{k}_3')\right)}$$

According to (1.1.6), (1.1.7) and (1.1.8), let's write combinations of random variables $\xi$ and respective wave vector combinations, giving non-zero contribution:

1. $\overline{\xi(\vec{k}_1)\xi(\vec{k}_1')}$: $\vec{k}_1 = -\vec{k}_1'$;
2. $\overline{\xi(\vec{k}_2)\xi(\vec{k}_2')\xi(\vec{k}_3)\xi(\vec{k}_3')}$:
   a) $\vec{k}_2 = -\vec{k}_2'$ & $\vec{k}_3 = -\vec{k}_3'$,
   b) $\vec{k}_2 = -\vec{k}_3'$ & $\vec{k}_3 = -\vec{k}_2'$,
   c) $\vec{k}_2 = -\vec{k}_3$ & $\vec{k}_2' = -\vec{k}_3'$;
3. $\overline{\xi(\vec{k}_2)\xi(\vec{k}_3)}$: $\vec{k}_2 = -\vec{k}_3$;
4. $\overline{\xi(\vec{k}_2')\xi(\vec{k}_3')}$: $\vec{k}_2' = -\vec{k}_3'$.

The terms, corresponding to variants 2c, 3 and 4 are constant. As we are interested in a variable component only, these combinations (along with the other constant terms) will be omitted. Hence we derive (intersections of the variants correspond to sets of zero measure):
$$C(\vec{r}) = (2\overline{S})^2\cdot\int F^2(\vec{k})e^{i\vec{k}\vec{r}}d\vec{k} + 2\left(\int F^2(\vec{k})e^{i\vec{k}\vec{r}}d\vec{k}\right)^2. \qquad (1.5.2)$$

The respective power spectrum is equal to
$$\mathcal{F}^2(\vec{k}) = (2\overline{S})^2 F^2(\vec{k}) + 2F_q^2(\vec{k}), \qquad (1.5.3)$$
$$F_q^2(\vec{k}) \equiv \int F^2(\vec{k}')F^2(\vec{k}-\vec{k}')d\vec{k}'. \qquad (1.5.4)$$

So the source function power spectrum can be expressed through a sum of two terms, which may be referred to as *linear* and *quadratic*. The linear term is proportional to the power spectrum of the original random process, the quadratic one — to the auto-convolution of this spectrum.

Let us find an approximation for the quadratic term in a partial case when an original random process has a power-law spectrum with the known outer scale and Kolmogorov spectral index:
$$F^2(k) = \frac{F_0^2}{k^{11/3}} \cdot e^{-\frac{k_0^2}{k^2}}. \qquad (1.5.5)$$

Analyzing the structure of the expression (1.5.4) with the account for (1.5.5) we can assume, that $F_q^2(\vec{k})$ must have the following asymptotics:

$$F_q^2(k) \approx F_0^4 J_0, \qquad k \ll k_0,$$

$$F_q^2(k) \approx 2F_0^4 \frac{J_1}{k^{11/3}}, \qquad k \gg k_0,$$

where

$$J_0 \equiv \frac{4\pi}{F_0^4} \int_0^\infty F^4(k)k^2 dk,$$

$$J_1 \equiv \frac{4\pi}{F_0^2} \int_0^\infty F^2(k)k^2 dk.$$

Now we can choose an approximation that fits the asymptotic conditions. A numerical evaluation shows, that the following approximating function is adequate enough:

$$F_q^2(k) \approx F_0^4 \frac{J_0}{\left(\left(\frac{J_0}{2J_1}\right)^{6/11} k^2 + 1\right)^{11/6}}.$$

Having found the values of $J_0$ and $J_1$, we obtain:

$$F_q^2(k) \approx \frac{F_{q0}^2}{\left(a\frac{k^2}{k_0^2}+1\right)^{11/6}}, \qquad (1.5.6)$$

where

$$F_{q0}^2 = \frac{1.514 F_0^4}{k_0^{13/3}}, \quad a = 0.1842.$$

So for the angular spectrum of the quadratic term we have:

$$C_q^2(l) = \frac{8\pi^3}{Rl} \cdot \int_0^\infty e^{-2\frac{l^2}{u^2}} F_q^2\left(\frac{u}{R}\right) du = \frac{4\sqrt{2}\pi^3 F_{q0}^2}{2^{11/6} R \cdot \left(\frac{alL}{2\pi R}\right)^{11/3}} \int_0^\infty \frac{e^{-t} t^{1/3} dt}{\left(\frac{1}{2}\left(\frac{2\pi R}{alL}\right)^2 t + 1\right)^{11/6}}$$

($k_0 = \frac{2\pi}{L}$, $L$ is outer scale).

We can find an approximation for the integral in the latter expression:

$$\int_0^\infty \frac{e^{-t} t^{1/3} dt}{\left(\frac{1}{2N^2} t + 1\right)^{11/6}} \approx \left(3.21 \cdot e^{-3.34 \cdot N} + 1\right) \cdot \int_0^\infty e^{-t\left(1 + \frac{11}{6} \frac{1}{2N^2}\right)} t^{1/3} dt = \left(3.21 \cdot e^{-3.34 \cdot N} + 1\right) \cdot \frac{\Gamma\left(\frac{4}{3}\right)}{\left(1 + \frac{11}{12N^2}\right)^{4/3}}.$$

So the angular spectrum of the quadratic term takes the following form:

$$C_q^2(l) = 1.48 \cdot 10^3 \cdot \frac{F_0^4 \cdot R^{8/3}}{\left(\frac{2\pi}{L}\right)^{2/3}} \cdot \frac{1 + 3.21 \cdot e^{-1.43\frac{l \cdot L}{2\pi R}}}{l^{11/3} \cdot (1 + 5.0 \cdot (\frac{2\pi R}{l \cdot L})^2)^{4/3}}. \qquad (1.5.7)$$

This spectrum behaves analogously to the linear variant with the difference, that its spectral index changes near the point $l_0 \approx 3 \cdot (2\pi R/L)$, instead of $l_0 = 2\pi R/L$ in the previous case, and its value is slightly bigger than unity in the region of lower $l$'s. The saturation when $R \to \infty$ also takes place. These effects, as in the linear case, are explained by the divergence of the primary spectrum (1.5.6) from the pure power law.

Now we can write a full angular spectrum corresponding to the given source function:

$$C_l^2 = 1.76 \cdot 10^2 \cdot \frac{\overline{S}^2 \cdot F_0^2 \cdot R^{8/3}}{l^{11/3} \cdot (1 + \frac{1}{2}(\frac{2\pi R}{l \cdot L})^2)^{4/3}} + 2.96 \cdot 10^3 \cdot \frac{F_0^4 \cdot R^{8/3}}{\left(\frac{2\pi}{L}\right)^{2/3}} \cdot \frac{1 + 3.21 \cdot e^{-1.43\frac{l \cdot L}{2\pi R}}}{l^{11/3} \cdot (1 + 5.0 \cdot (\frac{2\pi R}{l \cdot L})^2)^{4/3}}.$$

(1.5.8)

## 2. Galactic synchrotron emission

### 2.1. Source function

The source function of galactic synchrotron emission is proportional to $B_\perp^{(\gamma+1)/2}$ [10, 11, 12], where $B_\perp$ is a magnetic field projection to the picture plane, $\gamma$ is an electron energy spectrum index. The latter is bound with the radio frequency intensity spectral index $\alpha$ by expression $\gamma = 2\alpha + 1$. For the Galaxy $\alpha \approx 1$, and in this case the source function is proportional to the square of the perpendicular projection of the magnetic field.

On the other hand, from the theory of interstellar turbulence [2, 19] it is known, that Kolmogorov spectral index corresponds, in particular, to the quantity

$$\int d\Omega \sum_{l=1}^{3} G_{ll}(\vec{k}) \propto k^{-11/3},$$

where

$$G_{lm}(\vec{k}) \equiv \frac{1}{(2\pi)^3} \int B_{lm}(\vec{r}) e^{-i\vec{k}\vec{r}} d\vec{r},$$

$$B_{lm}(\vec{r}) \equiv \overline{b_l(\vec{r}\,') b_m(\vec{r}\,' + \vec{r})},$$

$\vec{b} \equiv \frac{1}{\sqrt{4\pi\rho}} \cdot \vec{B}$, $\rho$ is a medium density. If we suggest $\rho \approx const$, then

$$B_{lm}(\vec{r}) \approx \frac{1}{4\pi\rho} \overline{B_l(\vec{r}\,') B_m(\vec{r}\,' + \vec{r})}.$$

In a case of isotropic turbulence $B_{ll}(\vec{r})$ does not depend on index $l$ and direction of $\vec{r}$ [2]. In this case we have:

$$F_B^2(k) \propto k^{-11/3},$$

where $F_B^2(k)$ is a spatial power spectrum of a magnetic field component:

$$\overline{B_l(\vec{r}\,') B_l(\vec{r}\,' + \vec{r})} = \int F_B^2(k) e^{i\vec{k}\vec{r}} d\vec{k}.$$

Hence, if our suggestions are correct, a magnetic field component has Kolmogorov spectrum. And, then, if to assume that the orthogonal components of the picture plane projection of magnetic field are statistically independent, we can state that the synchrotron emission source function is proportional to square of a Kolmogorov process (with zero mean).

### 2.2. Angular spectrum

So, applying (1.5.7), we have the following expression for the synchrotron emission angular spectrum:

$$C_{syn}^2(\lambda, l) = A \cdot \lambda^{5.8} \cdot \alpha^{8/3} \frac{1 + 3.21 \cdot e^{-1.43 \frac{l}{2\pi\alpha q}}}{l^{11/3} \cdot (1 + 5.0 \cdot (\frac{2\pi\alpha q}{l})^2)^{4/3}}. \tag{2.2.1}$$

Here $\alpha \equiv R/R_P \approx 1/\sin|b|$, $R_P$ is a distance to a radiating region border in the direction of Galactic pole, $q \equiv R_P/L$, $\lambda$ is a wavelength in cm. Parameter $\alpha$ is introduced in order to account for a position of an observation site respectively to Galactic plane.

If we have two observational data sets corresponding to non-coinciding angular frequency bands, we can estimate two unknown parameters of this spectrum, $A$ and $q$. In this case, accounting for (A.8), we can write the following expression for a signal mean square for data sets with number $i = 1, 2$:

$$D_i = A \cdot \int \Phi_i^2(\vec{\kappa}) F^2(\lambda_i, \alpha_i, q, \kappa) d\vec{\kappa}, \tag{2.2.2}$$

where

$$F^2(\lambda,\alpha,q,\kappa) = \frac{\lambda^{5.8}\alpha^{8/3}}{(2\pi)^2} \cdot \frac{1 + 3.21 \cdot e^{-1.43\frac{\kappa}{2\pi\alpha q}}}{\kappa^{11/3} \cdot (1 + 5.0 \cdot (\frac{2\pi\alpha q}{\kappa})^2)^{4/3}},$$

$\Phi_i(\vec{\kappa})$ is an angular filter.

Let us rewrite (2.2.2) in the following form:

$$A = D_i \cdot f_i(q), \tag{2.2.3}$$

where

$$f_i(q) \equiv \left( \int \Phi_i^2(\vec{\kappa}) F^2(\lambda_i, \alpha_i, q, \kappa) d\vec{\kappa} \right)^{-1}.$$

So every data set yields to the respective dependence $A(q)$, and their intersection point gives an estimation of values of $A$ and $q$.

We used here two one-dimensional data sets corresponding to wavelengths 7.6cm ([24], see Figure 1) and 21cm ([9], see Figure 2).

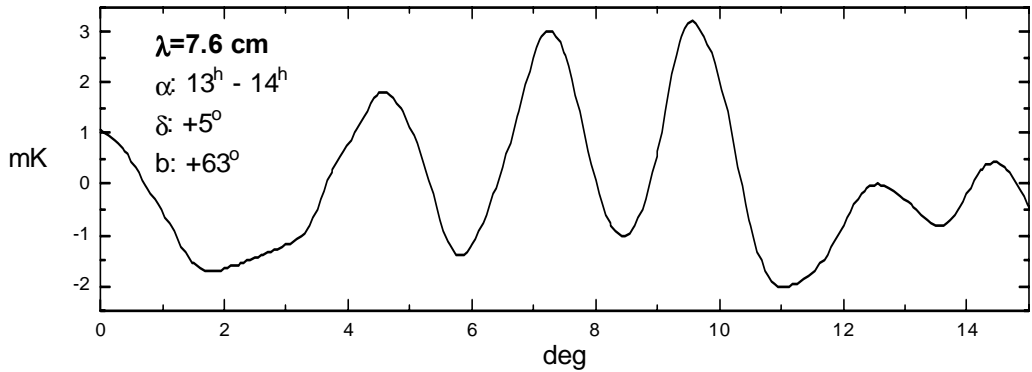

Figure 1. The experiment Cold-80 data (RATAN-600 radio telescope). The angular scale range $1° \div 5°$ is determined by data processing [24].

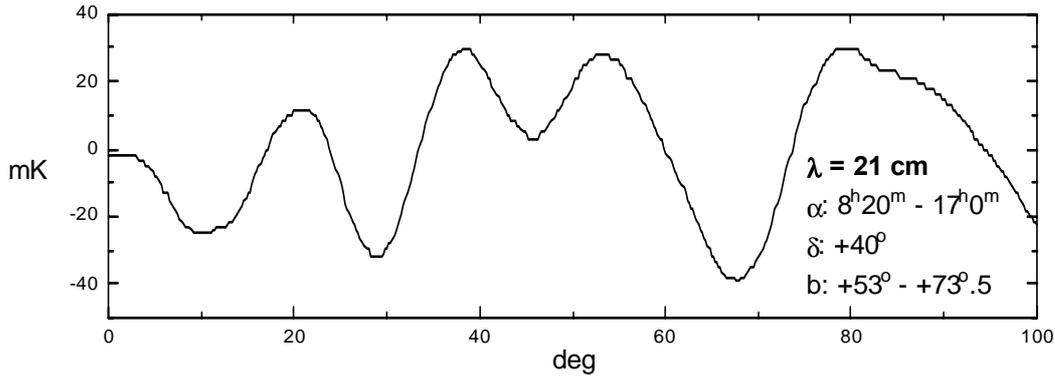

Figure 2. The 1420GHz survey data convoluted with a triple beam (beam separation $\pm 8°$, HPBW=$5°$). Discrete sources are previously removed [9].

For the RATAN-600 data we accounted for aperture efficiency $\varepsilon_a = 0.78$ and for all non-synchrotron components (discrete sources, free-free emission, dust emission) obtained by computer simulation [23, 4], which overall r.m.s. is estimated as $0.47\,mK$. Finally we obtained:

$\lambda_1 = 7.6\,cm$, $D_1 = 1.72 \cdot 10^{-6}\,K^2$, $\alpha_1 = 1.12$,

$$\Phi_1(\vec{\kappa}) = \left(1 - e^{-\frac{\kappa_x^2}{4a_1^2}}\right) e^{-\frac{\kappa_x^2}{4a_0^2}}, \text{ where } a_0 = \frac{2\sqrt{\ln 2}}{0.017}, \ a_1 = \frac{2\sqrt{\ln 2}}{0.087}.$$

For the second data set the respective parameters are:
$\lambda_2 = 21\, cm$, $D_2 = 3.88 \cdot 10^{-4}\, K^2$, $\alpha_2 = 1.13$,

$$\Phi_2(\vec{\kappa}) = (1 - \cos\kappa_x d) e^{-\frac{\kappa_x^2}{4a^2}}, \text{ where } a = \frac{2\sqrt{\ln 2}}{0.087}, \ d = 0.140.$$

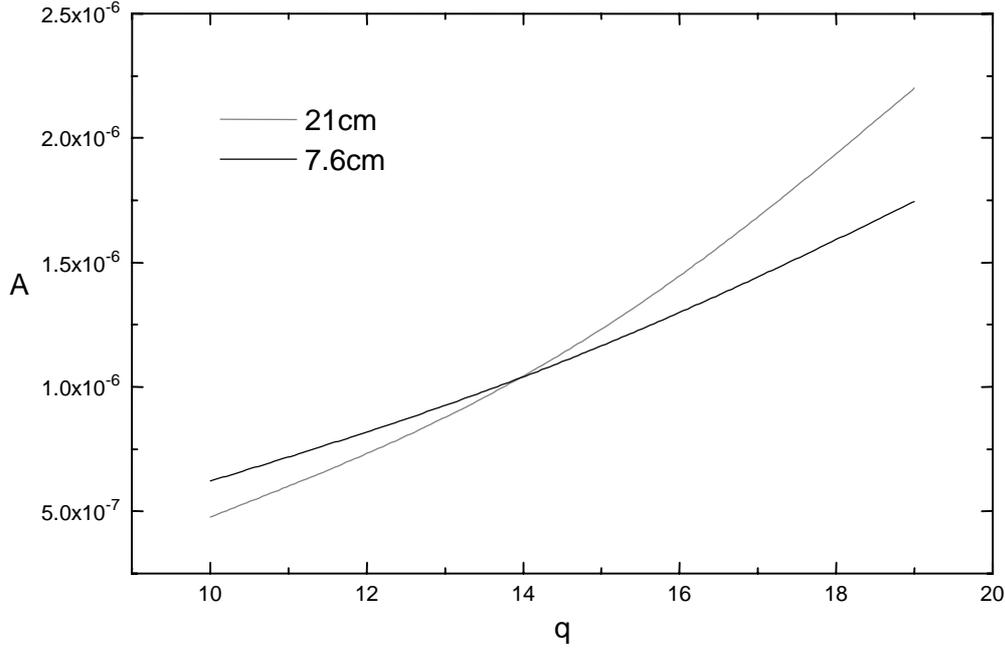

Figure 3. $A(q)$ functions corresponding to the considered data sets.

Now an intersection point can be found (see Figure 3) and it gives
$A = 1.04 \cdot 10^{-6}$, $q = 14.0$. (2.2.4)

### 2.3. A confidence interval for R/L

It is clearly seen from Figure 4 that small deviations of $D_i$ can cause a significant change of $A$ and $q$. So it is highly advisable to find a confidence interval of estimated $q$ and an accuracy of corresponding spectrum (2.2.1).

Using (B.3), we can estimate standard deviations of $D_i$:
$\sigma_{D_1} = 5.3 \cdot 10^{-7}\, K^2$, $\sigma_{D_2} = 1.1 \cdot 10^{-4}\, K^2$.

To simplify calculations we assume that $D_i$ values have normal distribution. In this case mutual probability distribution for $D_i$ has the following form ($D_i$ are statistically independent because the data sets correspond to different sky regions):

$$p_D(D_1, D_2) = \frac{1}{2\pi \sigma_{D_1} \sigma_{D_2}} e^{-\frac{(D_1 - \overline{D_1})^2}{2\sigma_{D_1}^2}} e^{-\frac{(D_2 - \overline{D_2})^2}{2\sigma_{D_2}^2}}.$$

Respective elementary probability is

$$dP = p_D(D_1, D_2)dD_1dD_2.$$

For area of an image of square $dD_1dD_2$ on a plane $\{(q,A)\}$, projected by transform (2.2.3), we can write the following:

$$dS = \frac{f_1(q)f_2(q)}{|D_1 f_1'(q) - D_2 f_2'(q)|}dD_1dD_2.$$

Accounting for (2.2.3), we can find mutual probability density of $A$ and $q$:

$$p(q,A) = \frac{dP}{dS} = A \cdot \frac{\left|\frac{f_1'(q)}{f_1(q)} - \frac{f_2'(q)}{f_2(q)}\right|}{f_1(q)f_2(q)} \cdot p_D\left(\frac{A}{f_1(q)}, \frac{A}{f_2(q)}\right). \quad (2.3.1)$$

Integrating it numerically over $A$, we obtain a probability density distribution of $q$ (see Figure 4).

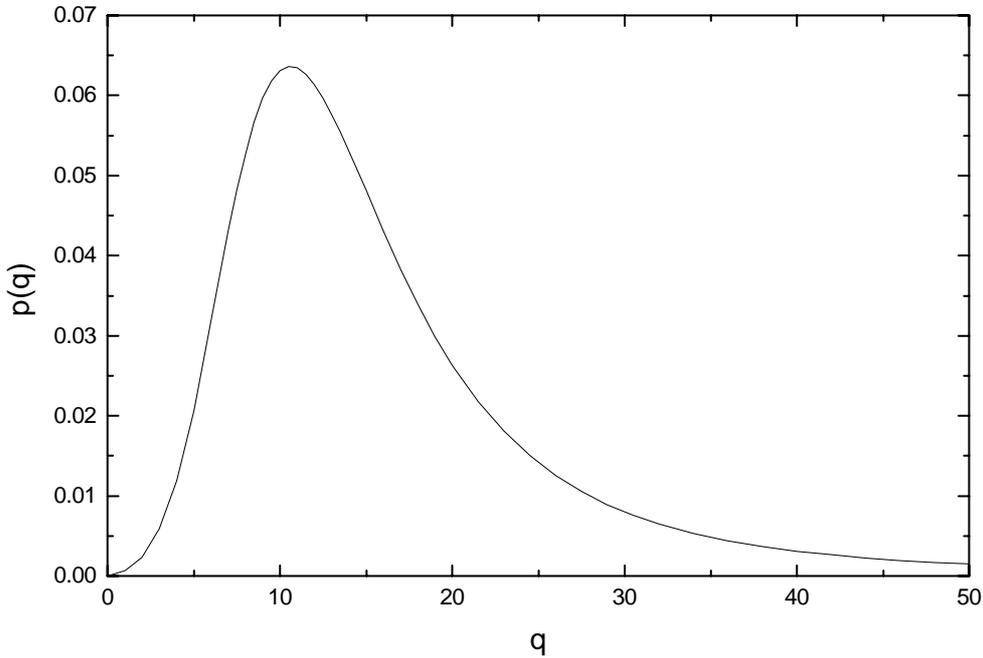

Figure 4. Probability density of the parameter $q = R/L$.

So we can estimate a confidence interval now. Finally we have:

$$R/L = 14.0^{+11.8}_{-5.7} \quad (2.3.2)$$

for confidence probability 0.68.

This error is not caused by the measurements accuracy and can be reduced only by observational areas enlarging.

Corresponding spectrum error is $\pm 16\%$ for $l = 100$, $\pm 39\%$ for $l = 320$ and $\pm 48\%$ for $l = 1000$.

### 2.4. The result

An angular spectrum, calculated according to (2.2.1) and (2.2.4) is shown on Figure 5 along with the empirical spectrum from the COBRAS/SAMBA project [3], obtained from 408 MHz and 1420MHz surveys data.

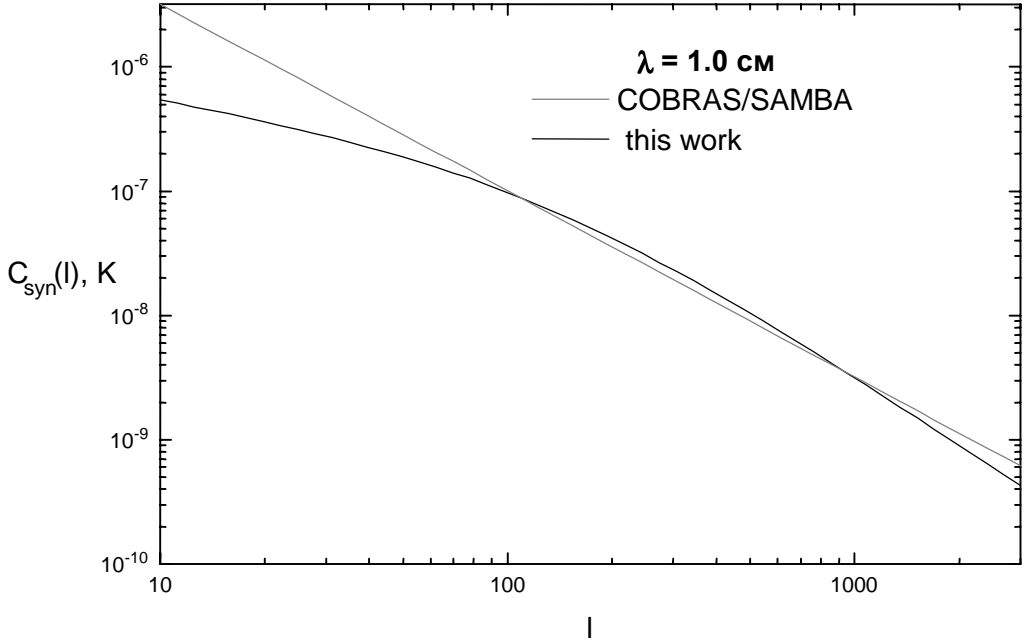

Figure 5. Angular spectrum of the Galactic synchrotron emission.

We can suggest, that the spectral index 3 found in [3], is caused by the influence of the outer scale $L$ and corresponds to transition from spectral index about unity at low $l$'s to Kolmogorov index $11/3$ when $l \to \infty$.

## 3. Galactic free-free emission

### 3.1. Source function

In the case of free-free emission source function is proportional to the squared electron density [20]. Respective brightness temperature is expressed through the line-of-sight integral from the electron density in the following way:

$$T_{ff} = 4.8 \cdot 10^{-6} \cdot \lambda^{2.16} \cdot \int_{l.o.s.} n_e^2 dr, \qquad (3.1.1)$$

where $\lambda$ is measured in $cm$, $n_e$ — in $cm^{-3}$, $r$ — in $pc$, $T_{ff}$ — in $K$.

On the other hand, it is known from pulsar emission scintillation measurements, that electron density fluctuations have Kolmogorov power spectrum:

$$F_e^2(k) = \frac{F_0^2}{k^{-11/3}},$$

where $k \geq 2\pi/L$, $L$ is an outer scale.

The amplitude of the spectrum is estimated in [7]:

$F_0^2 = 3.16 \cdot 10^{-4} \; m^{-20/3}$,

if $k$ is measured in $m^{-1}$.

So for the Galactic free-free emission angular spectrum we can use the result from section 1.5.

## 3.2. Angular spectrum

Applying (1.5.8) to (3.1.1) we have the following expression for the free-free emission angular spectrum:

$$C_{ff}^2(\lambda,l) = 4.0\cdot 10^{-10} \cdot \lambda^{4.32} \cdot \frac{\overline{n_e}^2 \cdot F_0^2 \cdot R^{8/3}}{l^{11/3}\cdot(1+0.5\cdot(\frac{2\pi R}{l\cdot L})^2)^{4/3}} + \\ + 6.65\cdot 10^{-10} \cdot \lambda^{4.32} \cdot \frac{F_0^4 \cdot R^{8/3}}{\left(\frac{2\pi}{L}\right)^{2/3}} \cdot \frac{1+3.21\cdot e^{-1.43\frac{l\cdot L}{2\pi R}}}{l^{11/3}\cdot(1+5.0\cdot(\frac{2\pi R}{l\cdot L})^2)^{4/3}} \qquad (3.2.1)$$

where

$\overline{n_e} = 0.05\,cm^{-3}$ is a mean electron density,

$F_0^2 = 3.16\cdot 10^{-4}\,m^{-20/3}$ is an electron density spatial power spectrum amplitude,

$R$ and $L$ are measured in $pc$, $\lambda$ in $cm$, the result — in $K^2$.

For $R$ one can assume $R = 1000/\sin|b|\ pc$, where $b$ is Galactic latitude [7].

The only remaining unknown parameter is $L$, the outer scale of turbulence. To estimate it we can use a signal r.m.s. value over some observation site. Using (A.7) we have the following expression for this value over circular region with radius $a$:

$$\sigma^2 = \frac{1}{2\pi}\int_0^\infty \left(1-\left(\frac{2}{ak}J_1(ak)\right)^2\right) C_{ff}^2(\lambda,k)\,k\,dk. \qquad (3.2.2)$$

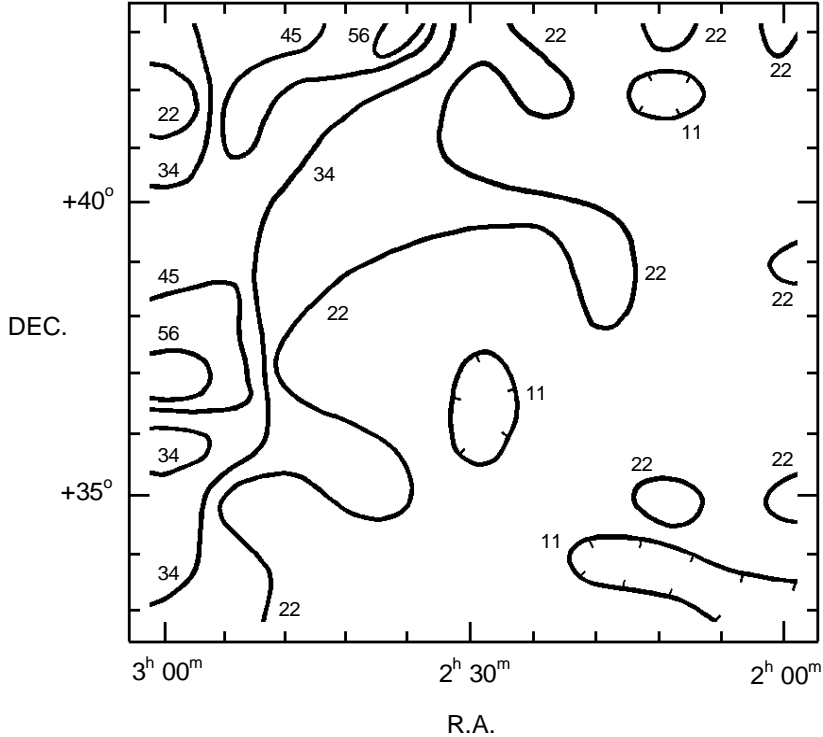

Figure 6. Brightness temperature of Galactic free-free emission in $\mu K$ at $\lambda = 1cm$ obtained by recounting from $H_\alpha$ intensity [27].

If $\sigma^2$ is known, one can fit the parameter $L$ of the spectrum (3.2.1) to satisfy (3.2.2). An observational site shape does not affect much (for non-pathological shapes), for another shape one can choose $a$ in order to make areas equal. To determine the outer scale $L$ we used $H_\alpha$ data, recounted to the free-free emission (see Figure 1, [27]). For $\sigma = 11.2\ \mu K$, $b = -21°$, $a = 6°.2$ fitting with (3.2.2) gives

$$L = 214\ pc.\hspace{4em}(3.2.3)$$

Free-free emission angular spectrum given by (3.2.1) with account for (3.2.3) is displayed on Figure 7 along with the spectrum, obtained in [3]. We can see, that this spectrum is $5 \div 10$ times lower than COBRAS/SAMBA estimation. The slope correspondent to spectral index 3 from [3] is observed only in the region $30 < l < 200$, for larger $l$ it corresponds to Kolmogorov spectral index.

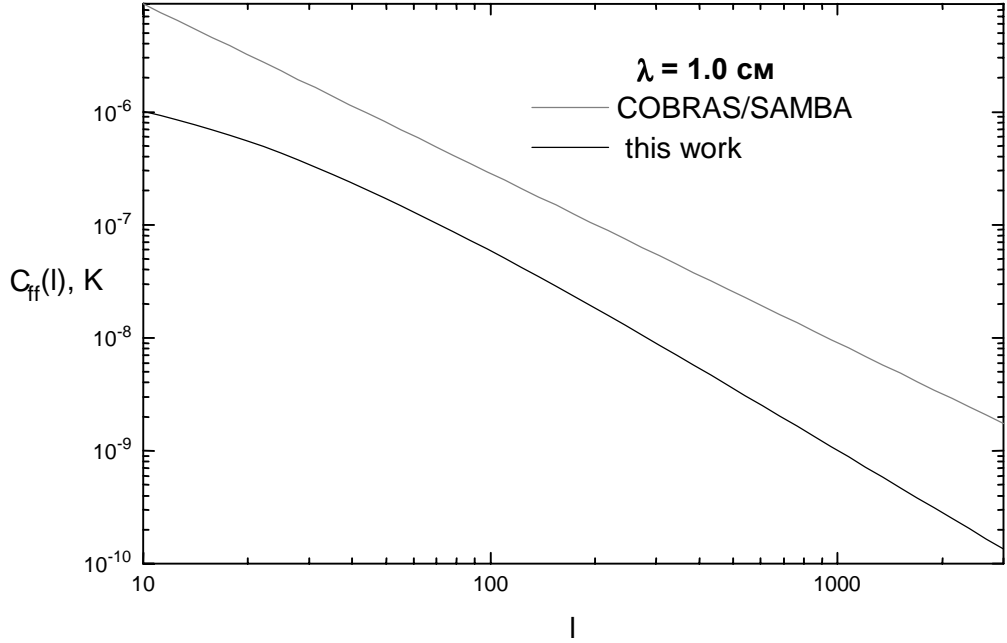

Figure 7. The Galactic free-free emission angular spectrum.

## 4. Summary

So the main results of the present work are as follows:
- ♦ A transform converting 3-d spatial spectrum of the source function to the angular spectrum of the observed picture is found in its general form, (1.2.3). Also a convenient approximate expression for this transform is found for Gaussian weighting function ((1.3.2), (1.3.4)).
- ♦ Angular spectra of the synchrotron and free-free Galactic radio emission are obtained with some general enough assumptions about their source functions ((2.2.1), (3.2.1)).
- ♦ Free parameters of these spectra are found by fitting to observational data. An information about turbulence outer scale is found for random magnetic field (2.3.2) and electron density (3.2.3).

However, we must denote the following disadvantages of the obtained results:
- ◊ As the behavior of interstellar medium parameters spatial spectra at the scales bigger than the turbulence outer scale is still not estimated, a cut–off such as the one used in (1.4.3) does not seem to be well-founded. So the outer scales found here may be regarded only as the order–of–magnitude estimations.
- ◊ Fitting of the synchrotron angular spectrum involves only a small part of the existing observational data, and may be regarded a demonstration of the method rather than a final result.

◊ Last estimations show, that $H_\alpha$ fluctuations are mostly caused by geo corona structure, and if it is true, then the free-free emission angular spectrum found here may be considered only as an upper limit, along with the correspondent outer scale.

So we can state now, that free-free emission has the angular spectrum lying at least one order of magnitude lower than present estimation of the COBRAS/SAMBA project, what shifts to lower radio frequencies the optimal band of CMB anisotropy experiments, and makes it possible to measure CMB angular spectrum even at short centimeter wavelengths.

## Acknowledgments

The author wishes to thank Acad. Yu.N. Parijskij, Prof. V.K. Dubrovich and Prof. P.A. Fridman for useful discussions. This work was partially supported by the Russian Fund for Fundamental Research, grant No.96-0216597 and by "COSMION".

## Appendix A. Estimation of a dispersion over a selected region

Let $s(\vec{r})$ to be a real random process with a spectral representation

$$s(\vec{r}) = \int e^{i\vec{k}\vec{r}} F(\vec{k}) \xi(\vec{k}) \sqrt{d\vec{k}} .$$

Let $\Omega$ to be a limited region in the space $\{\vec{r}\}$.
Then

$$D_\Omega \equiv \left\langle \left(s - \langle s \rangle_\Omega\right)^2 \right\rangle_\Omega = \tfrac{1}{\Omega}\int_\Omega \left( s(\vec{r}) - \tfrac{1}{\Omega}\int_\Omega s(\vec{r}')\,d\vec{r}' \right)^2 d\vec{r} = $$
$$= \int F(\vec{k}) \xi(\vec{k}) \sqrt{d\vec{k}} \cdot \int F(\vec{k}') \xi(\vec{k}') \sqrt{d\vec{k}'} \cdot \left( \Pi_\Omega(\vec{k}+\vec{k}') - \Pi_\Omega(\vec{k}) \Pi_\Omega(\vec{k}') \right) \quad \text{(A.1)}$$

where

$$\Pi_\Omega(\vec{k}) \equiv \tfrac{1}{\Omega} \int_\Omega e^{i\vec{k}\vec{r}} d\vec{r} . \quad \text{(A.2)}$$

We assume below that region $\Omega$ is symmetrical with respect to coordinate inversion (in this case $\Pi_\Omega(\vec{k})$ is a real function) and contains $\vec{0}$.

With account for this assumption we can write some other properties of $\Pi_\Omega(\vec{k})$:

$$\Pi_\Omega(\vec{0}) = 1, \quad \text{(A.3)}$$
$$\Pi_\Omega(-\vec{k}) = \Pi_\Omega(\vec{k}), \quad \text{(A.4)}$$
$$\int \Pi_\Omega(\vec{k}) d\vec{k} = \tfrac{(2\pi)^n}{\Omega}, \quad \text{(A.5)}$$
$$\int \Pi_\Omega^2(\vec{k}) d\vec{k} = \tfrac{(2\pi)^n}{\Omega}. \quad \text{(A.6)}$$

where $n$ is a dimension of $\{\vec{r}\}$.

For rough estimations it is convenient to assume that $\Pi_\Omega(\vec{k})$ is equal to unity in some vicinity of $\vec{0}$ of volume $\tfrac{(2\pi)^n}{\Omega}$, and is equal to zero in all other points.

Having averaged (A.1) with account for (1.1.6), (1.1.7), (1.1.8), (A.3) and (A.4) we have:

$$\overline{D_\Omega} = \int F^2(\vec{k})\left(1 - \Pi_\Omega^2(\vec{k})\right) d\vec{k} . \quad \text{(A.7)}$$

If we can neglect harmonics with a characteristic scale larger than the dimension of $\Omega$, the second multiplicand in the latter expression can be omitted:

$$\overline{D_\Omega} \approx \int F^2(\vec{k}) d\vec{k} . \quad \text{(A.8)}$$

## Appendix B. Standard deviation of an estimated dispersion

With the account for (A.1), we can write the following:

$$\sigma_D^2 \equiv \overline{\left(D_\Omega - \overline{D_\Omega}\right)^2} = \overline{D_\Omega^2} - \overline{D_\Omega}^2 =$$
$$= \int F(\vec{k}) \sqrt{d\vec{k}} \cdot \int F(\vec{k}') \sqrt{d\vec{k}'} \cdot \int F(\vec{k}_1) \sqrt{d\vec{k}_1} \cdot \int F(\vec{k}_1') \sqrt{d\vec{k}_1'} \cdot$$
$$\cdot \left(\Pi_\Omega(\vec{k}+\vec{k}') - \Pi_\Omega(\vec{k})\Pi_\Omega(\vec{k}')\right)\left(\Pi_\Omega(\vec{k}_1+\vec{k}_1') - \Pi_\Omega(\vec{k}_1)\Pi_\Omega(\vec{k}_1')\right) \cdot$$
$$\cdot \overline{\xi(\vec{k})\xi(\vec{k}')\xi(\vec{k}_1)\xi(\vec{k}_1')} - \overline{D_\Omega}^2$$

Then, accounting for (1.1.6), (1.1.7) and (1.1.8), we can write wave vector combinations giving a non-zero contribution to $\overline{D_\Omega^2}$:

1) $\vec{k} = -\vec{k}'$ & $\vec{k}_1 = -\vec{k}_1'$,
2) $\vec{k} = -\vec{k}_1$ & $\vec{k}' = -\vec{k}_1'$,

3) $\vec{k} = -\vec{k}_1' \ \& \ \vec{k}' = -\vec{k}_1$.

Variant 1) yields to $\overline{D_\Omega}^2$ (see (A.7)) and cancels, combinations 2) and 3) give an equal result. So we have:

$$\sigma_D^2 = 2\int F^2(\vec{k})d\vec{k} \cdot \int F^2(\vec{k}')d\vec{k}' \cdot \left(\Pi_\Omega(\vec{k}+\vec{k}') - \Pi_\Omega(\vec{k})\Pi_\Omega(\vec{k}')\right)^2. \qquad (B.1)$$

And, again, if harmonics with a characteristic scale larger than the dimension of $\Omega$ are negligible, the latter expression can be simplified. As $F(\vec{k})$ is even, we have:

$$\sigma_D^2 \approx 2\int F^2(\vec{k})d\vec{k} \cdot \int F^2(\vec{k}')d\vec{k}' \cdot \Pi_\Omega^2(\vec{k}-\vec{k}'). \qquad (B.2)$$

If we suppose that the region where $\Pi_\Omega^2(\vec{k}-\vec{k}')$ with fixed $\vec{k}'$ is significantly non-zero is small enough to neglect spectrum variations in it, then with account for (A.6) we have the following approximate expression:

$$\sigma_D^2 \approx 2\frac{(2\pi)^n}{\Omega}\int F^4(\vec{k})d\vec{k}. \qquad (B.3)$$